\documentclass[12pt]{article}
\usepackage{amsfonts}
\usepackage{mathrsfs}
\usepackage{amsmath}
\usepackage{verbatim}
\usepackage{amssymb}
\usepackage{amsfonts,amsmath,latexsym,amssymb,txfonts,bm}
\usepackage[american]{babel}

\def\II{{\mathbb I}}
\def\RR{{\mathbb R}}

\def\be{\begin{equation}}
\def\ee{\end{equation}}
\def\bea{\begin{eqnarray}}
\def\eea{\end{eqnarray}}

\def\sideremark#1{\ifvmode\leavevmode\fi\vadjust{\vbox to0pt{\vss
\hbox to 0pt{\hskip\hsize\hskip1em
\vbox{\hsize2cm\tiny\raggedright\pretolerance10000 \noindent
#1\hfill}\hss}\vbox to8pt{\vfil}\vss}}}

\begin{document}

\begin{titlepage}
\thispagestyle{empty} \null

\hspace*{50truemm}{\hrulefill}\par\vskip-4truemm\par
\hspace*{50truemm}{\hrulefill}\par\vskip5mm\par
\hspace*{50truemm}{{\large\sc New Mexico Tech {\rm
(\today)}}}\vskip4mm\par
\hspace*{50truemm}{\hrulefill}\par\vskip-4truemm\par
\hspace*{50truemm}{\hrulefill}

\vspace{3cm} \centerline{\huge\bf Noncommutative Einstein
Equations}
\bigskip
\centerline{\huge\bf}
\bigskip
\bigskip
\centerline{\Large\bf Guglielmo Fucci and Ivan G. Avramidi}
\bigskip
\centerline{\it New Mexico Institute of Mining and Technology}
\centerline{\it Socorro, NM 87801, USA} \centerline{\it E-mail:
iavramid@nmt.edu, gfucci@nmt.edu}
\bigskip
\medskip
\vfill

{\narrower
\par
We study a noncommutative deformation of general relativity where
the gravitational field is described by a matrix-valued symmetric
two-tensor field. The equations of motion are derived in the
framework of this new theory by varying a diffeomorphisms and
gauge invariant action constructed by using a matrix-valued scalar
curvature. Interestingly the genuine noncommutative part of the
dynamical equations is described only in terms of a particular
tensor density that vanishes identically in the commutative limit.
A noncommutative generalization of the energy-momentum tensor for
the matter field is studied as well.
\par}

\end{titlepage}

\section{Introduction}

The basic notions of General Relativity (GR) are closely related
with the geometric interpretation of second order elliptic partial
differential operators describing the propagation of massless
scalar fields without internal structure \cite{AV1}. Let us
consider a manifold $\textit{M}$ without boundary. Let $f\in
C^{\infty}(\textit{M})$ be a smooth function on $\textit{M}$ and
let $L:C^{\infty}(\textit{M})\longrightarrow
C^{\infty}(\textit{M})$ be a second order partial differential
operator of the form
\begin{equation}\label{1}
L=-a^{\mu\nu}(x)\partial_{\mu}\partial_{\nu}
+b^{\mu}(x)\partial_{\mu}+c(x)\;,
\end{equation}
with smooth real coefficients $a^{\mu\nu}$, $b^\mu$ and $c$. In
particular, $a^{\mu\nu}$ is a real smooth symmetric non-degenerate
matrix at each point of the manifold $\textit{M}$, that is, for
any point $x$ in $\textit{M}$
\begin{eqnarray}
a^{\mu\nu}(x)=a^{\nu\mu}(x)\;,\nonumber \qquad \det
a^{\mu\nu}(x)\neq 0\;. \nonumber
\end{eqnarray}
Since the matrix $a^{\mu\nu}$ is symmetric, it has real
eigenvalues and since the determinant is different from zero
through the whole manifold $\textit{M}$, it has constant
signature. If the number of negative eigenvalues is $1$, the
operator $L$ is called \emph{hyperbolic}. Hyperbolic operators
defined on $\textit{M}$ describe the propagation of massless
scalar fields, in particular photons, over the manifold
$\textit{M}$.

It can be easily shown that for the operator $L$ to be covariant
the matrix $a^{\mu\nu}$ should transform like a tensor of type
$(2,0)$ under diffeomorphisms of the manifold $\textit{M}$. This
enables us to identify the matrix $a^{\mu\nu}$ with a Riemannian
metric $g^{\mu\nu}$ over $\textit{M}$. Once a metric $g^{\mu\nu}$
is defined one can construct all the geometrical quantities needed
to develop the theory of General Relativity. In this theory the
gravitational field is described by a symmetric non-degenerate
tensor of type $(0,2)$ and its dynamics is described by Einstein's
equations.

A natural generalization of the GR theory leads us to consider the
gravitational field described by a $(N\times N)$ \emph{matrix} of
metrics $(g^{\mu\nu})^{A}{}_{B}$ instead of a single metric (for
details see \cite{AV1,AV2}). Such a matrix-valued metric appear
naturally if one considers the propagation of a field with an
internal structure (a multiplet of fields rather than a single
field). In this case the coefficients of the operator L, that is,
the functions $a^{\mu\nu}(x)$, $b^{\mu}(x)$ and $c(x)$, become
matrix-valued. This follows from the idea that at short distances,
or at high energies, the role played by the photon can be played
by a multiplet of gauge fields \cite{AV1}.

It is important to stress here that this approach for deforming
general relativity is different from the ones proposed in the
framework of noncommutative geometry \cite{FU,KS,MM,HR}, where the
coordinates do not commute and the standard product between
functions is replaced by the Moyal product \cite{Ma}.

The purpose of this paper is to derive the equations of motion for
the field $a^{\mu\nu}$ that generalizes the role played by
$g^{\mu\nu}$ in the general theory of relativity. Since this model
is a noncommutative extension of Einstein's general relativity we
will call the corresponding equations of motions noncommutative
Einstein's equations.

The outline of the paper is as follows. First we develop the
necessary formalism that will be used in the derivation of the
dynamical equations of the theory. More precisely we will describe
a generalization of various important geometrical object to
endomorphism-valued quantities. Then we construct an action that
is invariant under diffeomorphisms of the manifold $M$ and gauge
transformations and contains the matrix-valued scalar curvature.
By varying, then, the action with respect to the new dynamical
field $a^{\mu\nu}$, we obtain the equations of motion.

\section{Matrix Geometry}

We will describe, in this section, the formalism needed in order
to write the action for matrix general relativity following
\cite{AV1,AV2} and derive the equations of motion of the theory.
This formalism is a generalization of differential geometry, which
is the natural language of general relativity, and it will be
called \emph{Matrix Geometry}. The formalism that we are going to
describe is related to the algebra-valued formulation of Mann
\cite{Ma1} and Wald \cite{Wa1}.

As already stated before, the main idea is to describe the
gravitational field as a matrix-valued symmetric two-tensor field.
Let $M$ be an $n$-dimensional Riemannian manifold without boundary
and let $g^{\mu\nu}$ be a metric tensor defined on the tangent
bundle $TM$. It is well known that General Relativity is nothing
but the dynamical theory of the metric 2-tensor field which is,
basically, an isomorphism between tangent and cotangent bundles.
In our model the metric 2-tensor field is replaced by a
endomorphism-valued 2-tensor field $a^{\mu\nu}$ which represents
an isomorphism of more general bundles over the manifold $M$. The
main idea here, similar to general relativity, is to develop a
dynamical theory of this endomorphism-valued 2-tensor field
$a^{\mu\nu}$. This generality brings a much richer structure and
content to the model.

Let $V$ be an $N$-dimensional Hermitian vector bundle over $M$,
let $\mathscr{T}=TM\otimes V$ be the bundle constructed by taking
the tensor product of the tangent bundle to the manifold $M$ with
the vector bundle $V$, and let
$\mathscr{T^{\ast}}=T^{\ast}M\otimes V$, where $T^{\ast}M$ is the
cotangent bundle to $M$. We consider, then, the bundle
$\textrm{Iso}(\mathscr{T},\mathscr{T^{\ast}})$, elements of which,
are isomorphisms between $\mathscr{T}$ and $\mathscr{T^{\ast}}$.
In general the isomorphisms
$\textit{B}:\mathscr{T}\longrightarrow\mathscr{T^{\ast}}$ can be
identified with the sections of the bundle $T^{\ast}M\otimes
T^{\ast}M\otimes\textrm{End}(V)$ and the isomorphisms
$\textit{A}:\mathscr{T^{\ast}}\longrightarrow\mathscr{T}$ with the
sections of the bundle $TM\otimes TM\otimes\textrm{End}(V)$.

Let $a$ be a symmetric self-adjoint element of $TM\otimes
TM\otimes\textrm{End}(V)$, that is, \be
a^{\mu\nu}=a^{\nu\mu},\qquad (a^{\mu\nu})^*=a^{\mu\nu}\,. \ee This
element is an isomorphism between the bundles $\mathscr{T^{\ast}}$
and $\mathscr{T}$ if the following equation
\begin{equation}\label{2}
a^{\mu\nu}\phi_{\nu}=\psi^{\mu}\;,
\end{equation}
with arbitrary $\psi\in\mathscr{T^{\ast}}$, has a unique solution
\begin{equation}\label{3}
\phi_{\nu}=b_{\nu\mu}\psi^{\mu}\;.
\end{equation}
This last requirement can be cast in another form; the element $a$
is an isomorphism if the following equation
\begin{equation}\label{4}
a^{\mu\nu}b_{\nu\rho}=b_{\mu\nu}a^{\nu\rho}=\delta_{\mu}^{\rho}\cdot\mathbb{I}\;,
\end{equation}
has a unique solution $b\in T^{\ast}M\otimes
T^{\ast}M\otimes\textrm{End}(V)$.

There are some properties of the matrix $b_{\mu\nu}$ that need
attention. The first property is the following: the matrix
$b_{\mu\nu}$ satisfies the following equation
\begin{equation}\label{5}
b_{\mu\nu}^{\ast}=b_{\nu\mu}\;,
\end{equation}
but it is not necessarily a self-adjoint matrix symmetric in its
tensor indices. Moreover, one can use $a^{\mu\nu}$ and
$b_{\mu\nu}$ to lower and raise indices, although particular care
is required in this operations because, in general, $a^{\mu\nu}$
and $b_{\mu\nu}$ do not commute and $b_{\mu\nu}$ is not symmetric
in its tensorial indices \cite{AV1}.

We need, now, to develop some kind of invariant calculus. For this
purpose, we introduce matrix-valued Christoffel symbols
$\mathscr{A}^{\mu}{}_{\alpha\beta}$ that transform, under
diffeomorphisms of $M$, as connection coefficients, namely
\begin{equation}\label{6}
\mathscr{A}^{\prime\mu^{\prime}}{}_{\alpha^{\prime}\beta^{\prime}}(x^{\prime})=\frac{\partial
x^{\prime\mu}}{\partial x^{\nu}}\frac{\partial
x^{\gamma}}{\partial x^{\prime\alpha}}\frac{\partial
x^{\delta}}{\partial
x^{\prime\beta}}\mathscr{A}^{\nu}{}_{\gamma\delta}(x)+\frac{\partial
x^{\prime\mu}}{\partial
x^{\nu}}\frac{\partial^{2}x^{\nu}}{\partial
x^{\prime\alpha}\partial x^{\prime\beta}}\cdot\mathbb{I}\;,
\end{equation}
where $\II$ is the identity matrix. Once we have a connection on
$M$ we can define a way to differentiate tensors.

Let $T_{p}^{q}(M)$ be the tensor bundle of type $(p,q)$ on the
manifold $M$. We define the following linear map
\begin{equation}\label{7}
\mathcal{D}:T_{p}^{q}\otimes V\longrightarrow T_{p}^{q+1}\otimes V
\end{equation}
by
\begin{equation}\label{8}
\mathcal{D}_\alpha\phi_{\nu_{1}\ldots\nu_{q}}^{\mu_{1}\ldots\mu_{p}}
=\partial_{\alpha}\phi^{\mu_{1}\ldots\mu_{p}}_{\nu_{1}\ldots\nu_{q}}
+\sum_{j=1}^{p}\mathscr{A}^{\mu_{j}}{}_{\lambda\alpha}\phi^{\mu_{1}
\ldots\mu_{j-1}\lambda\mu_{j+1}\ldots\mu_{p}}_{\nu_{1}\ldots
\nu_{q}}-\sum_{i=1}^{q}\mathscr{A}^{\lambda}{}_{\nu_{i}\alpha}
\phi^{\mu_{1}\ldots\mu_{p}}_{\nu_{1}\ldots\nu_{i-1}
\lambda\nu_{i+1}\ldots\nu_{q}}\;.
\end{equation}
This map is a well defined operator between $T_{p}^{q}\otimes V$
and $T_{p}^{q+1}\otimes V$. It is important to stress, at this
point, that the linear map defined in (\ref{8}) is not a covariant
differentiation. In fact, it is easy to show that, because of the
non-commutativity, the Leibnitz rule does not hold. However, in
the commutative limit, the operator (\ref{7}) reduces to an
ordinary covariant derivative on the manifold $M$.

In complete analogy with the ordinary Riemannian geometry, we
shall define the matrix curvature and the matrix torsion tensors.
Let $\phi\in T^{\ast}M\otimes V$, we compute
\begin{equation}\label{9}
\mathcal{D}_{\mu}\mathcal{D}_{\nu}\phi_{\alpha}
-\mathcal{D}_{\nu}\mathcal{D}_{\mu}\phi_{\alpha}
=-\mathcal{R}^{\lambda}{}_{\alpha\mu\nu}\phi_{\lambda}
+\mathcal{T}^{\lambda}{}_{\mu\nu}
\mathcal{D}_{\lambda}\phi_{\alpha}\;,
\end{equation}
where
\begin{equation}\label{10}
\mathcal{R}^{\lambda}{}_{\alpha\mu\nu}
=\partial_{\mu}\mathscr{A}^{\lambda}{}_{\alpha\nu}
-\partial_{\nu}\mathscr{A}^{\lambda}{}_{\alpha\mu}
+\mathscr{A}^{\lambda}{}_{\beta\mu}\mathscr{A}^{\beta}{}_{\alpha\nu}
-\mathscr{A}^{\lambda}{}_{\beta\nu}\mathscr{A}^{\beta}{}_{\alpha\mu}\;,
\end{equation}
and
\begin{equation}\label{11}
\mathcal{T}^{\lambda}{}_{\mu\nu}=\mathscr{A}^{\lambda}{}_{\mu\nu}-\mathscr{A}^{\lambda}{}_{\nu\mu}\;.
\end{equation}
Once the matrix curvature (Riemann) tensor is defined we can
construct the matrix Ricci tensor, namely
\begin{equation}\label{12}
\mathcal{R}_{\mu\nu}=\mathcal{R}^{\alpha}{}_{\mu\alpha\nu}\;.
\end{equation}
In order to write the action for matrix gravity, we need to
introduce the matrix scalar curvature $\mathcal{R}$. Since the
metric $a^{\mu\nu}$ and the Ricci tensor $\mathcal{R}_{\mu\nu}$
are matrices, they do not commute in general and the definition of
the scalar curvature, obtained by contracting the metric tensor
with the Ricci tensor from the left, would be different if the
contraction would be performed with the metric tensor on the
right. In order to avoid this choice, we use a symmetrized
definition of the matrix-valued scalar curvature as follows
\begin{equation}\label{13}
\mathcal{R}=\frac{1}{2}\left(a^{\mu\nu}\mathcal{R}_{\mu\nu}+\mathcal{R}_{\mu\nu}a^{\mu\nu}\right)\;.
\end{equation}

We need, now, to relate the connection coefficients
$\mathscr{A}^{\rho}{}_{\mu\nu}$ to the metric tensor $a^{\mu\nu}$.
We impose a compatibility condition similar to the one in
Riemannian geometry as
\begin{equation}\label{14}
\partial_{\mu}a^{\alpha\beta}+\mathscr{A}^{\alpha}{}_{\lambda\mu}a^{\lambda\beta}+\mathscr{A}^{\beta}{}_{\lambda\mu}a^{\alpha\lambda}=0\;.
\end{equation}
The above equation has the following solution in a closed form
\cite{AV1}
\begin{equation}\label{15}
\mathscr{A}^{\alpha}{}_{\lambda\mu}=\frac{1}{2}b_{\lambda\sigma}(a^{\alpha\gamma}\partial_{\gamma}a^{\rho\sigma}-a^{\rho\gamma}\partial_{\gamma}a^{\sigma\alpha}-a^{\sigma\gamma}\partial_{\gamma}a^{\alpha\rho}+S^{\alpha\rho\sigma}+S^{\rho\sigma\alpha}+S^{\sigma\rho\alpha})b_{\rho\mu}\;,
\end{equation}
where $S$ is an arbitrary matrix-valued tensor that satisfies the
symmetry property
\begin{equation}\label{16}
S^{\alpha\mu\nu}=-S^{\alpha\nu\mu}\;.
\end{equation}
The matrix-valued tensor $S$ is related, in the general case, with
the torsion $\mathcal{T}$ in (\ref{11}). Moreover, in the
commutative limit, the tensor $S$ reduces exactly to the torsion.
It is important to notice that in matrix geometry the connection
(\ref{15}) is not symmetric in the two lower indices even if the
tensor $S$ vanishes. In the rest of the paper we will assume,
without loss of generality, that the tensor $S$ vanishes.

In order to write an action for the model under consideration we
need a generalization of the concept of measure. As a guiding
principle, any generalization of the measure $\mu$ has to lead, in
the commutative limit, to the ordinary Riemannian measure
$\sqrt{|\det g_{\mu\nu}|}$. Moreover the measure $\mu$ is a
density depending only on the metric $a^{\mu\nu}$ and not on its
derivatives and transforming, under diffeomorphisms of $M$, as
\begin{equation}\label{17}
\mu^{\prime}(x^{\prime})=J(x)\mu(x)\;,
\end{equation}
where
\begin{equation}\label{18}
J(x)=\det\left(\frac{\partial x^{\prime\mu}(x)}{\partial
x^{\alpha}}\right).
\end{equation}
A definition of the measure $\mu$ which is a straightforward
generalization of the Riemannian measure, is the following
\cite{AV1}
\begin{equation}\label{19}
\mu=\frac{1}{N}\textrm{Tr}_{\;V}\rho\;,
\end{equation}
where $\rho$ is a matrix-valued scalar density, which can be
defined, for example, as follows
\begin{equation}\label{20}
\rho=\int_{\mathbb{R}^{n}}\frac{d\xi}{\pi^{\frac{n}{2}}}
\exp(-a^{\mu\nu}\xi_{\mu}\xi_{\nu})\;.
\end{equation}
Then $\rho$ only depends on the metric $a$ and transforms in the
correct way under diffeomorphisms of $M$.

\section{Variation of the Action}

We construct the action functional for the field $a^{\mu\nu}$
following \cite{AV1}. This functional has to be invariant under
both diffeomorphisms of $M$ and gauge transformations. The
infinitesimal form of these transformations is
\begin{equation}\label{21}
\delta_{\omega}a^{\mu\nu}=[\omega,a^{\mu\nu}]\;,
\end{equation}
and
\begin{equation}\label{22}
\delta_{\xi}a^{\mu\nu}=\mathscr{L}_{\xi}a^{\mu\nu}\;,
\end{equation}
where $\omega$ is an element of the algebra of the gauge group
\begin{displaymath}
\delta_{\omega}U=\omega\;,
\end{displaymath}
and $\xi$ is the generator of the infinitesimal coordinate
transformation
\begin{displaymath}
\delta_{\xi}x^{\mu}=\mathscr{L}_{\xi}x^{\mu}=-\xi^{\mu}\;.
\end{displaymath}

We can construct an action functional for field $a^{\mu\nu}$ that
satisfies the properties described above, by using the
matrix-valued scalar curvature, defined in (\ref{13}), and the
measure (\ref{19}). A good candidate for the action is the
following
\begin{equation}\label{23}
S_{\textrm{MGR}}(a)=\frac{1}{16\pi
G}\int_{M}dx\frac{1}{N}\textrm{Tr}_{\;V}(\rho\;\mathcal{R})\;.
\end{equation}
It is worth noticing that because of the cyclic property of the
trace, the relative position of $\rho$ and the scalar curvature is
irrelevant, moreover it is easily shown that the action functional
(\ref{23}) is invariant under the diffeomorphisms (\ref{22}) and
under the gauge transformations (\ref{21}). Since the action has
the invariant properties discussed above, the currents associated
with the symmetries satisfy the identities
\begin{equation}\label{23a}
\partial_\alpha \left( a^{\mu\alpha} \frac{\delta S}{\delta a^{\mu\lambda}} \right)
+\frac{1}{2} ( \partial_\lambda a^{\mu\nu} ) \frac{\delta
S}{\delta a^{\mu\nu}} = 0\;,
\end{equation}
and
\begin{equation}\label{23b}
\left[ a^{\mu\nu} , \frac{\delta S}{\delta a^{\mu\nu}}\right]=0\;,
\end{equation}
where the (\ref{23a}) is the current generated by the invariance
with respect to diffeomorphisms and (\ref{23b}) is the current
generated by the internal (gauge) symmetry. The above identities
represent an endomorphism-valued generalization of the No\"{e}ther
identities, which are related, in the usual theory, to the
contracted Bianchi identities.

Now that we have the action functional we can derive the equations
of motion for the field $a^{\mu\nu}$, which is the {\it main goal
and the main result of the present paper}. These equations will be
matrix-valued and they will constitute a generalization of the
ordinary Einstein's equations that we will call {\it
noncommutative Einstein equations}. In order to find the dynamics
of the model we vary the action (\ref{23}) with respect to the
field $a^{\mu\nu}$ considered as independent variable, namely
\begin{displaymath}
a^{\mu\nu}\longrightarrow a^{\mu\nu}+\delta a^{\mu\nu}\;.
\end{displaymath}
By doing so we obtain, for the variation of the action, the
following
\begin{equation}\label{24}
\delta S=S(a^{\mu\nu}+\delta
a^{\mu\nu})-S(a^{\mu\nu})=\frac{1}{16\pi
G}\int_{M}dx\frac{1}{N}\textrm{Tr}_{\;V}(\mathcal{G}_{\mu\nu}\delta
a^{\mu\nu})\;,
\end{equation}
where $\mathcal{G}_{\mu\nu}$ is some matrix valued symmetric
tensor density. Then, of course, the desired equations of motion
are
\begin{equation}\label{25}
\mathcal{G}_{\mu\nu}=0\;.
\end{equation}
It is important to notice that the matrix-valued tensor density
(\ref{25}) has to coincide with the Einstein tensor in the
commutative limit, more precisely we need that, in the commutative
limit, the following relation holds
\begin{equation}\label{26}
\sqrt{g}\left(R_{\mu\nu}-\frac{1}{2}g_{\mu\nu}R\right)=
\frac{1}{N}\textrm{Tr}\;\mathcal{G}_{\mu\nu}\;.
\end{equation}

Our main task, then, is to find the explicit form of the equations
of motion that result from the variation of the action (\ref{23}).
In all the calculations that will follow the {\it order of the
terms is important}, unless explicitly stated, due to the matrix
nature of them.

First of all, we rewrite the action in a more explicit form which
is more suitable for the subsequent variation, namely
\begin{equation}\label{27}
S_{\textrm{MGR}}(a)=\frac{1}{16\pi
G}\int_{M}dx\frac{1}{N}\textrm{Tr}_{\;V}
\left[\rho\frac{1}{2}(a^{\mu\nu}\mathcal{R}_{\mu\nu}
+\mathcal{R}_{\mu\nu}a^{\mu\nu})\right]\;.
\end{equation}
By varying the terms in (\ref{27}) with respect to the independent
field $a^{\mu\nu}$, and by using the cyclic property of the trace
we get
\begin{eqnarray}\label{28}
\delta S_{\textrm{MGR}}(a)=\frac{1}{16\pi
G}\int_{M}dx\frac{1}{N}\textrm{Tr}_{\;V}\left[\delta\rho\mathcal{R}+\frac{1}{2}\{\mathcal{R}_{\mu\nu},\rho\}\delta
a^{\mu\nu}+\frac{1}{2}\{\rho,
a^{\mu\nu}\}\delta\mathcal{R}_{\mu\nu}\right]\;,
\end{eqnarray}
where the curly brackets $\{\;,\;\}$ denote anti-commutation,
namely $\{A,B\}=AB+BA$.

From the expressions (\ref{10}) and (\ref{12}), we can evaluate
the variation of the matrix-valued Ricci tensor, more precisely we
have
\begin{eqnarray}\label{29}
\delta\mathcal{R}_{\mu\nu}&=&\partial_{\alpha}(\delta\mathscr{A}^{\alpha}{}_{\mu\nu})-\partial_{\nu}(\delta\mathscr{A}^{\alpha}{}_{\mu\alpha})+\delta\mathscr{A}^{\alpha}{}_{\lambda\alpha}\mathscr{A}^{\lambda}{}_{\mu\nu}+\mathscr{A}^{\alpha}{}_{\lambda\alpha}\delta\mathscr{A}^{\lambda}{}_{\mu\nu}+\nonumber\\
&-&\delta\mathscr{A}^{\alpha}{}_{\lambda\nu}\mathscr{A}^{\lambda}{}_{\mu\alpha}-\mathscr{A}^{\alpha}{}_{\lambda\nu}\delta\mathscr{A}^{\lambda}{}_{\mu\alpha}\;.
\end{eqnarray}
From now on, for simplicity of notation, we set \be
B^{\mu\nu}\equiv\{\rho,a^{\mu\nu}\}\,. \ee By substituting
(\ref{29}) in (\ref{28}), and by using the cyclic property of the
trace we obtain
\begin{eqnarray}\label{30}
\delta S_{\textrm{MGR}}(a)&=&\frac{1}{16\pi
G}\int_{M}dx\frac{1}{N}\textrm{Tr}_{\;V}\bigg[\delta\rho\mathcal{R}+\frac{1}{2}\{\mathcal{R}_{\mu\nu},\rho\}\delta
a^{\mu\nu}+\frac{1}{2}B^{\mu\nu}\partial_{\alpha}(\delta\mathscr{A}^{\alpha}{}_{\mu\nu})+\nonumber\\
&-&\frac{1}{2}B^{\mu\nu}\partial_{\nu}(\delta\mathscr{A}^{\alpha}{}_{\mu\alpha})+\frac{1}{2}\mathscr{A}^{\lambda}{}_{\mu\nu}B^{\mu\nu}\delta\mathscr{A}^{\alpha}{}_{\lambda\alpha}+\frac{1}{2}B^{\mu\nu}\mathscr{A}^{\alpha}{}_{\lambda\alpha}\delta\mathscr{A}^{\lambda}{}_{\mu\nu}+\nonumber\\
&-&\frac{1}{2}\mathscr{A}^{\lambda}{}_{\mu\alpha}B^{\mu\nu}\delta\mathscr{A}^{\alpha}{}_{\lambda\nu}-\frac{1}{2}B^{\mu\nu}\mathscr{A}^{\alpha}{}_{\lambda\nu}\delta\mathscr{A}^{\lambda}{}_{\mu\alpha}\bigg]\;.
\end{eqnarray}
By integrating by parts and by collecting similar terms we get
\begin{eqnarray}\label{31}
\delta S_{\textrm{MGR}}(a)&=&\frac{1}{16\pi
G}\int_{M}dx\frac{1}{N}\textrm{Tr}_{\;V}\bigg[\delta\rho\mathcal{R}+\frac{1}{2}\{\mathcal{R}_{\mu\nu},\rho\}\delta
a^{\mu\nu}-\frac{1}{2}\bigg(B^{\mu\nu}{}_{,\alpha}-B^{\mu\nu}\mathscr{A}^{\lambda}{}_{\alpha\lambda}+\nonumber\\
&+&\mathscr{A}^{\mu}{}_{\lambda\alpha}B^{\lambda\nu}+B^{\mu\lambda}\mathscr{A}^{\nu}{}_{\alpha\lambda}\bigg)\delta\mathscr{A}^{\alpha}{}_{\mu\nu}+\frac{1}{2}\bigg(B^{\mu\nu}{}_{,\nu}+\mathscr{A}^{\mu}{}_{\lambda\nu}B^{\lambda\nu}\bigg)\delta\mathscr{A}^{\alpha}{}_{\mu\alpha}\bigg]\;.
\end{eqnarray}
We can rewrite the last expression in a more compact form, namely
\begin{equation}\label{32}
\delta S_{\textrm{MGR}}(a)=\frac{1}{16\pi
G}\int_{M}dx\frac{1}{N}\textrm{Tr}_{\;V}\bigg[\delta\rho\mathcal{R}+\frac{1}{2}\{\mathcal{R}_{\mu\nu},\rho\}\delta
a^{\mu\nu}-\frac{1}{2}C^{\mu\nu}{}_{\alpha}\delta\mathscr{A}^{\alpha}{}_{\mu\nu}+\frac{1}{2}D^{\mu}\delta\mathscr{A}^{\alpha}{}_{\mu\alpha}\bigg]\;,
\end{equation}
where the matrix-valued tensor densities $C^{\mu\nu}{}_{\alpha}$
and $D^{\mu}$ have the explicit expression
\begin{eqnarray}\label{33}
C^{\mu\nu}{}_{\alpha}&=&\{a^{\mu\nu},\rho_{,\alpha}-\rho\mathscr{A}^{\lambda}{}_{\alpha\lambda}\}-\rho[a^{\mu\nu},\mathscr{A}^{\lambda}{}_{\alpha\lambda}]-[\rho,\mathscr{A}^{\mu}{}_{\rho\alpha}]a^{\rho\nu}+\nonumber\\
&-&\{\rho,[\mathscr{A}^{\nu}{}_{\lambda\alpha},a^{\mu\lambda}]\}-a^{\mu\lambda}[\mathscr{A}^{\nu}{}_{\lambda\alpha},\rho]+2\{\rho,a^{\mu\lambda}\}\mathscr{A}^{\nu}{}_{[\alpha\lambda]}\;,
\end{eqnarray}
and
\begin{equation}\label{34}
D^{\mu}=\{a^{\mu\nu},\rho_{,\nu}-\mathscr{A}^{\rho}{}_{\nu\rho}\rho\}-[\rho,\mathscr{A}^{\mu}{}_{\rho\nu}]a^{\rho\nu}-[\rho,\mathscr{A}^{\rho}{}_{\nu\rho}]a^{\mu\nu}\;.
\end{equation}
It is worth noticing that in the commutative limit, or, in other
words, when all the matrices commute, the tensor densities
$C^{\mu\nu}{}_{\alpha}$ and $D^{\mu}$ are identically zero, and
the variation of the action $\delta S_{\textrm{MGR}}$ simply
reduces to the standard result of the general theory of
relativity.

We can write, now, the variation of the connection coefficients.
By using the expression (\ref{15}), and by noticing that
\begin{displaymath}
\delta b_{\mu\nu}=-b_{\mu\rho}(\delta
a^{\rho\sigma})b_{\sigma\nu}\;,
\end{displaymath}
we obtain the following
\begin{eqnarray}\label{35}
\delta\mathscr{A}^{\alpha}{}_{\lambda\mu}&=&-b_{\lambda\nu}\delta
a^{\nu\beta}\mathscr{A}^{\alpha}{}_{\beta\mu}-\mathscr{A}^{\alpha}{}_{\lambda\nu}\delta
a^{\nu\beta}b_{\beta\mu}+\frac{1}{2}b_{\lambda\sigma}\delta
a^{\alpha\gamma}(\partial_{\gamma}a^{\rho\sigma})b_{\rho\mu}+\nonumber\\
&-&\frac{1}{2}b_{\lambda\sigma}\delta
a^{\rho\gamma}(\partial_{\gamma}a^{\sigma\alpha})b_{\rho\mu}
-\frac{1}{2}b_{\lambda\sigma}\delta
a^{\sigma\gamma}(\partial_{\gamma}a^{\rho\alpha})b_{\rho\mu}+\frac{1}{2}b_{\lambda\sigma}a^{\alpha\gamma}(\partial_{\gamma}\delta
a^{\rho\sigma})b_{\rho\mu}+\nonumber\\
&-&\frac{1}{2}b_{\lambda\sigma}a^{\rho\gamma}(\partial_{\gamma}\delta
a^{\sigma\alpha})b_{\rho\mu}-\frac{1}{2}b_{\lambda\sigma}a^{\sigma\gamma}(\partial_{\gamma}\delta
a^{\rho\alpha})b_{\rho\mu}\;.
\end{eqnarray}
Once we have the explicit expression for the variation of the
connection coefficients, we can evaluate the last two terms that
appear in the variation of the action (\ref{32}). We start with
the first of the two
\begin{eqnarray}\label{36}
&-&\frac{1}{2}\int_{M}dx\;\textrm{Tr}_{\;V}(C^{\mu\nu}{}_{\alpha}\delta\mathscr{A}^{\alpha}{}_{\mu\nu})=\nonumber\\
&=&\frac{1}{2}\int_{M}dx\;\textrm{Tr}_{\;V}\bigg[\mathscr{A}^{\alpha}{}_{\beta\nu}C^{\mu\nu}{}_{\alpha}b_{\mu\lambda}\delta
a^{\lambda\beta}
+b_{\beta\nu}C^{\mu\nu}{}_{\alpha}\mathscr{A}^{\alpha}{}_{\mu\lambda}\delta
a^{\lambda\beta}+\nonumber\\
&-&\frac{1}{2}(\partial_{\gamma}a^{\rho\sigma})b_{\rho\nu}C^{\mu\nu}{}_{\alpha}b_{\mu\sigma}\delta
a^{\alpha\gamma}
+\frac{1}{2}(\partial_{\gamma}a^{\alpha\sigma})b_{\rho\nu}C^{\mu\nu}{}_{\alpha}b_{\mu\sigma}\delta
a^{\rho\gamma}+\nonumber\\
&+&\frac{1}{2}(\partial_{\gamma}a^{\rho\alpha})b_{\rho\nu}C^{\mu\nu}{}_{\alpha}b_{\mu\sigma}\delta
a^{\sigma\gamma}-\frac{1}{2}b_{\rho\nu}C^{\mu\nu}{}_{\alpha}b_{\mu\sigma}a^{\alpha\gamma}(\partial_{\gamma}\delta
a^{\rho\sigma})+\nonumber\\
&+&\frac{1}{2}b_{\rho\nu}C^{\mu\nu}{}_{\alpha}b_{\mu\sigma}a^{\rho\gamma}(\partial_{\gamma}\delta
a^{\alpha\sigma})+\frac{1}{2}b_{\rho\nu}C^{\mu\nu}{}_{\alpha}b_{\mu\sigma}a^{\sigma\gamma}(\partial_{\gamma}\delta
a^{\rho\alpha})\bigg]\;,
\end{eqnarray}
where in this last expression we used the cyclic property of the
trace.

We introduce the following definition, which will be useful in
order to simplify the notation,
\begin{equation}\label{37}
F_{\beta\alpha\rho}=b_{\beta\nu}C^{\mu\nu}{}_{\alpha}b_{\mu\rho}\;.
\end{equation}
By using the above definition, the expression in (\ref{36}) can be
rewritten as follows
\begin{eqnarray}\label{38}
&-&\frac{1}{2}\int_{M}dx\;\textrm{Tr}_{\;V}(C^{\mu\nu}{}_{\alpha}\delta\mathscr{A}^{\alpha}{}_{\mu\nu})=\nonumber\\
&=&\frac{1}{2}\int_{M}dx\;\textrm{Tr}_{\;V}\bigg[\mathscr{A}^{\alpha}{}_{\beta\gamma}a^{\gamma\rho}F_{\rho\alpha\lambda}\delta
a^{\lambda\beta}+F_{\beta\alpha\rho}a^{\rho\gamma}\mathscr{A}^{\alpha}{}_{\gamma\lambda}\delta
a^{\lambda\beta}+\nonumber\\
&-&\frac{1}{2}(\partial_{\gamma}a^{\sigma\rho})F_{\rho\alpha\sigma}\delta
a^{\alpha\gamma}+\frac{1}{2}(\partial_{\gamma}a^{\alpha\sigma})F_{\rho\alpha\sigma}\delta
a^{\rho\gamma}+\frac{1}{2}(\partial_{\gamma}a^{\alpha\rho})F_{\rho\alpha\sigma}\delta
a^{\sigma\gamma}+\nonumber\\
&-&\frac{1}{2}F_{\rho\alpha\sigma}a^{\alpha\gamma}(\partial_{\gamma}\delta
a^{\rho\sigma})+\frac{1}{2}F_{\rho\alpha\sigma}a^{\rho\gamma}(\partial_{\gamma}\delta
a^{\alpha\sigma})+\frac{1}{2}F_{\rho\alpha\sigma}a^{\sigma\gamma}(\partial_{\gamma}\delta
a^{\rho\alpha})\bigg]\;,
\end{eqnarray}
where the first two terms in the last expression has been derived
by using the relation
\begin{equation}\label{39}
\mathscr{A}^{\alpha}{}_{\beta\nu}C^{\mu\nu}{}_{\alpha}b_{\mu\lambda}\delta
a^{\lambda\beta}=\mathscr{A}^{\alpha}{}_{\beta\gamma}a^{\gamma\rho}b_{\rho\nu}C^{\mu\nu}{}_{\alpha}b_{\mu\lambda}\delta
a^{\lambda\beta}=\mathscr{A}^{\alpha}{}_{\beta\gamma}a^{\gamma\rho}F_{\rho\alpha\lambda}\delta
a^{\lambda\beta}\;.
\end{equation}
By integrating by parts and by relabelling dummy indices we find
the final expression for (\ref{38}), namely
\begin{eqnarray}\label{40}
-\frac{1}{2}\int_{M}dx\;\textrm{Tr}_{\;V}(C^{\mu\nu}{}_{\alpha}\delta\mathscr{A}^{\alpha}{}_{\mu\nu})&=&\frac{1}{2}\int_{M}dx\;\textrm{Tr}_{\;V}\bigg[\mathscr{A}^{\alpha}{}_{\beta\gamma}a^{\gamma\rho}F_{\rho\alpha\gamma}+F_{\beta\alpha\rho}a^{\rho\gamma}\mathscr{A}^{\alpha}{}_{\gamma\lambda}+\nonumber\\
&-&\frac{1}{2}(\partial_{\beta}a^{\rho\sigma})\bigg(F_{\rho\lambda\sigma}-F_{\lambda\sigma\rho}-F_{\sigma\rho\lambda}\bigg)\bigg]\delta
a^{\lambda\beta}+\nonumber\\
&+&\frac{1}{4}\bigg\{\partial_{\gamma}\bigg[\bigg(F_{\rho\lambda\sigma}-F_{\lambda\sigma\rho}-F_{\sigma\rho\lambda}\bigg)a^{\lambda\gamma}\bigg]\bigg\}\delta
a^{\rho\sigma}\;.
\end{eqnarray}

For the last term in the variation of the action (\ref{32}), we
use similar arguments which lead us to the expression (\ref{40}).
In this case we introduce the following definition:
\begin{equation}\label{41}
G_{\beta\alpha\rho}=b_{\beta\alpha}D^{\mu }b_{\mu\rho}\;.
\end{equation}
By using the definition above and the cyclic property of the trace
we obtain
\begin{eqnarray}\label{41a}
&\phantom{-}&\frac{1}{2}\int_{M}dx\;\textrm{Tr}_{\;V}(D^{\mu}\delta\mathscr{A}^{\alpha}{}_{\mu\alpha})=\nonumber\\
&-&\frac{1}{2}\int_{M}dx\;\textrm{Tr}_{\;V}\bigg[\mathscr{A}^{\alpha}{}_{\beta\rho}a^{\rho\gamma}G_{\gamma\alpha\lambda}\delta
a^{\lambda\beta}+G_{\beta\alpha\gamma}a^{\gamma\rho}\mathscr{A}^{\alpha}{}_{\rho\lambda}\delta
a^{\lambda\beta}+\nonumber\\
&-&\frac{1}{2}(\partial_{\gamma}a^{\rho\sigma})G_{\rho\alpha\sigma}\delta
a^{\alpha\gamma}+\frac{1}{2}(\partial_{\gamma}a^{\alpha\sigma})G_{\rho\alpha\sigma}\delta
a^{\rho\gamma}+\frac{1}{2}(\partial_{\gamma}a^{\rho\alpha})G_{\rho\alpha\sigma}\delta
a^{\sigma\gamma}+\nonumber\\
&-&\frac{1}{2}G_{\rho\alpha\sigma}a^{\alpha\gamma}(\partial_{\gamma}\delta
a^{\rho\sigma})+\frac{1}{2}G_{\rho\alpha\sigma}a^{\rho\gamma}(\partial_{\gamma}\delta
a^{\alpha\sigma})+\frac{1}{2}G_{\rho\alpha\sigma}a^{\sigma\gamma}(\partial_{\gamma}\delta
a^{\rho\alpha})\bigg]\;.
\end{eqnarray}
By integrating by parts and relabelling dummy indices we get
\begin{eqnarray}\label{42}
\frac{1}{2}\int_{M}dx\;\textrm{Tr}_{\;V}(D^{\mu}\delta\mathscr{A}^{\alpha}{}_{\mu\alpha})&=&-\frac{1}{2}\int_{M}dx\;\textrm{Tr}_{\;V}\bigg[\mathscr{A}^{\alpha}{}_{\beta\rho}a^{\rho\gamma}G_{\gamma\alpha\lambda}+G_{\beta\alpha\gamma}a^{\rho\gamma}\mathscr{A}^{\alpha}{}_{\rho\lambda}+\nonumber\\
&-&\frac{1}{2}(\partial_{\beta}a^{\rho\sigma})\bigg(G_{\rho\lambda\sigma}-G_{\lambda\sigma\rho}-G_{\sigma\rho\lambda}\bigg)\bigg]\delta
a^{\lambda\beta}+\nonumber\\
&-&\frac{1}{4}\bigg\{\partial_{\gamma}\bigg[\bigg(G_{\rho\lambda\sigma}-G_{\lambda\sigma\rho}-G_{\sigma\rho\lambda}\bigg)a^{\lambda\gamma}\bigg]\bigg\}\delta
a^{\rho\sigma}\;.
\end{eqnarray}

It is worth noticing that in the above expressions, (\ref{40}) and
(\ref{42}), the tensor densities $F$ and $G$ always appear in the
same combination. This observation justifies the following
definitions
\begin{equation}\label{43}
X_{\rho\lambda\sigma}=F_{\rho\lambda\sigma}-F_{\lambda\sigma\rho}-F_{\sigma\rho\lambda}\;,
\end{equation}
and
\begin{equation}\label{44}
Y_{\rho\lambda\sigma}=G_{\rho\lambda\sigma}-G_{\lambda\sigma\rho}-G_{\sigma\rho\lambda}\;.
\end{equation}
By using the two definitions above we can rewrite the arguments of
the traces in (\ref{40}) and in (\ref{42}) respectively as
\begin{eqnarray}\label{45}
-\frac{1}{2}C^{\mu\nu}{}_{\alpha}\delta\mathscr{A}^{\alpha}{}_{\mu\nu}&=&\frac{1}{2}\bigg[\mathscr{A}^{\alpha}{}_{\beta\gamma}a^{\gamma\rho}F_{\rho\alpha\lambda}+F_{\beta\alpha\rho}a^{\rho\gamma}\mathscr{A}^{\alpha}{}_{\gamma\lambda}-\frac{1}{2}(\partial_{\beta}a^{\rho\sigma})X_{\rho\lambda\sigma}+\nonumber\\
&+&\frac{1}{2}\partial_{\gamma}(X_{\lambda\rho\beta}a^{\rho\gamma})\bigg]\delta
a^{\lambda\beta}\;,
\end{eqnarray}
and
\begin{eqnarray}\label{46}
\frac{1}{2}D^{\mu}\delta\mathscr{A}^{\alpha}{}_{\mu\alpha}&=&-\frac{1}{2}\bigg[\mathscr{A}^{\alpha}{}_{\beta\rho}a^{\rho\gamma}G_{\gamma\alpha\lambda}+G_{\beta\alpha\gamma}a^{\rho\gamma}\mathscr{A}^{\alpha}{}_{\rho\lambda}-\frac{1}{2}(\partial_{\beta}a^{\rho\sigma})Y_{\rho\lambda\sigma}+\nonumber\\
&+&\frac{1}{2}\partial_{\gamma}(Y_{\lambda\rho\beta}a^{\rho\gamma})\bigg]\delta
a^{\lambda\beta}\;.
\end{eqnarray}
By combining the results (\ref{45}) and (\ref{46}) we obtain the
expression for the last two terms in the variation of the action,
namely
\begin{eqnarray}\label{47}
-\frac{1}{2}C^{\mu\nu}{}_{\alpha}\delta\mathscr{A}^{\alpha}{}_{\mu\nu}+\frac{1}{2}D^{\mu}\delta\mathscr{A}^{\alpha}{}_{\mu\alpha}=\frac{1}{2}\bigg\{\mathscr{A}^{\alpha}{}_{\beta\rho}a^{\rho\gamma}(F_{\gamma\alpha\lambda}-G_{\gamma\alpha\lambda})+\nonumber\\
+(F_{\beta\alpha\gamma}-G_{\beta\alpha\gamma})a^{\gamma\rho}\mathscr{A}^{\alpha}{}_{\rho\gamma}+\frac{1}{2}(\partial_{\beta}a^{\rho\sigma})(Y_{\rho\lambda\sigma}-X_{\rho\lambda\sigma})+\nonumber\\
-\frac{1}{2}\partial_{\gamma}\bigg[(Y_{\lambda\rho\beta}-X_{\lambda\rho\beta})a^{\rho\gamma}\bigg]\bigg\}\delta
a^{\lambda\beta}\;.
\end{eqnarray}

\section{Noncommutative Einstein Equations}

With the expression (\ref{47}) for the last two terms in
(\ref{32}), the variation of the action has the form (\ref{24})
which is suitable for the derivation of the dynamical equations of
the model. Before writing the complete dynamical equations, we
will simplify further the expression (\ref{47}).

The definition (\ref{43}) gives a linear relation between the
matrix-valued tensor density $X$ and a particular combination of
matrix-valued tensor density $F$, a similar linear relation
between $Y$ and $G$ is given in (\ref{44}). By using simple tensor
algebra, it can be easily shown that those relations can be
inverted, namely we can write
\begin{equation}\label{48}
F_{\rho\lambda\sigma}=-\frac{1}{2}(X_{\lambda\sigma\rho}+X_{\sigma\rho\lambda})\;,
\end{equation}
and
\begin{equation}\label{49}
G_{\rho\lambda\sigma}=-\frac{1}{2}(Y_{\lambda\sigma\rho}-Y_{\sigma\rho\lambda})\;.
\end{equation}
By substituting the equations (\ref{48}) and (\ref{49}) in the
expression (\ref{47}) we obtain the following
\begin{eqnarray}\label{50}
-\frac{1}{2}C^{\mu\nu}{}_{\alpha}\delta\mathscr{A}^{\alpha}{}_{\mu\nu}&+&\frac{1}{2}D^{\mu}\delta\mathscr{A}^{\alpha}{}_{\mu\alpha}=\frac{1}{4}\bigg\{\mathscr{A}^{\alpha}{}_{\beta\rho}a^{\rho\gamma}[(Y_{\alpha\lambda\gamma}-X_{\alpha\lambda\gamma})+(Y_{\lambda\gamma\alpha}-X_{\lambda\gamma\alpha})]+\nonumber\\
&+&[(Y_{\alpha\gamma\lambda}-X_{\alpha\gamma\lambda})+(Y_{\gamma\beta\alpha}-X_{\gamma\beta\alpha})]a^{\rho\gamma}\mathscr{A}^{\alpha}{}_{\rho\lambda}+\nonumber\\
&+&(\partial_{\beta}a^{\rho\sigma})(Y_{\rho\lambda\sigma}-X_{\rho\lambda\sigma})-\partial_{\gamma}[(Y_{\lambda\rho\beta}-X_{\lambda\rho\beta})a^{\rho\gamma}]\bigg\}\delta
a^{\lambda\beta}\;.
\end{eqnarray}

We can see, in the last formula, that the tensor densities $X$ and
$Y$ enter always in the same combination. It is useful, therefore,
to define the following tensor density
\begin{equation}\label{51}
H_{\mu\nu\rho}=Y_{\mu\nu\rho}-X_{\mu\nu\rho}\;.
\end{equation}
With this last definition we can rewrite (\ref{50}) as
\begin{eqnarray}\label{52}
-\frac{1}{2}C^{\mu\nu}{}_{\alpha}\delta\mathscr{A}^{\alpha}{}_{\mu\nu}+\frac{1}{2}D^{\mu}\delta\mathscr{A}^{\alpha}{}_{\mu\alpha}=\frac{1}{4}\bigg[\mathscr{A}^{\alpha}{}_{\beta\rho}a^{\rho\gamma}H_{\alpha\lambda\gamma}+\mathscr{A}^{\alpha}{}_{\beta\rho}a^{\rho\gamma}H_{\lambda\gamma\alpha}+\nonumber\\
+H_{\alpha\gamma\beta}a^{\rho\gamma}\mathscr{A}^{\alpha}{}_{\rho\lambda}+H_{\gamma\beta\alpha}a^{\rho\gamma}\mathscr{A}^{\alpha}{}_{\rho\lambda}+(\partial_{\beta}a^{\rho\sigma})H_{\rho\lambda\sigma}-\partial_{\gamma}(H_{\lambda\rho\beta}a^{\rho\gamma})\bigg]\delta
a^{\lambda\beta}\;.
\end{eqnarray}
By using the compatibility condition (\ref{14}), we can write that
\begin{equation}\label{53}
\partial_{\beta}a^{\rho\sigma}=-\mathscr{A}^{\rho}{}_{\gamma\beta}a^{\gamma\sigma}-\mathscr{A}^{\sigma}{}_{\gamma\beta}a^{\rho\gamma}\;,
\end{equation}
moreover we obtain that
\begin{equation}\label{54}
-\partial_{\gamma}(H_{\lambda\rho\beta}a^{\rho\gamma})=-(\partial_{\gamma}H_{\lambda\rho\beta})a^{\rho\gamma}+H_{\lambda\rho\beta}\mathscr{A}^{\rho}{}_{\sigma\gamma}a^{\sigma\gamma}+H_{\lambda\rho\beta}\mathscr{A}^{\gamma}{}_{\sigma\gamma}a^{\sigma\rho}\;.
\end{equation}
Since $H_{\mu\nu\rho}$ is a tensor density, we can write
\begin{equation}\label{55}
\mathcal{D}_{\gamma}H_{\lambda\rho\beta}=\partial_{\gamma}H_{\lambda\rho\beta}-\mathscr{A}^{\alpha}{}_{\lambda\gamma}H_{\alpha\rho\beta}-\mathscr{A}^{\alpha}{}_{\rho\gamma}H_{\lambda\alpha\beta}-\mathscr{A}^{\alpha}{}_{\beta\gamma}H_{\lambda\rho\alpha}-\mathscr{A}^{\alpha}{}_{\gamma\alpha}H_{\lambda\rho\beta}\;.
\end{equation}
By using the results obtained in (\ref{53}), (\ref{54}) and
(\ref{55}) we can express (\ref{52}) as follows
\begin{eqnarray}\label{56}
-\frac{1}{2}C^{\mu\nu}{}_{\alpha}\delta\mathscr{A}^{\alpha}{}_{\mu\nu}+\frac{1}{2}D^{\mu}\delta\mathscr{A}^{\alpha}{}_{\mu\alpha}=\frac{1}{4}\bigg\{2\mathscr{A}^{\alpha}{}_{[\beta\rho]}a^{\rho\gamma}H_{\alpha\lambda\gamma}+2H_{\alpha\lambda\beta}a^{\rho\gamma}\mathscr{A}^{\alpha}{}_{[\rho\lambda]}+\nonumber\\
-(\mathcal{D}_{\gamma}H_{\lambda\rho\beta})a^{\rho\gamma}-[\mathscr{A}^{\alpha}{}_{\lambda\rho},H_{\alpha\gamma\beta}]a^{\rho\gamma}-H_{\alpha\gamma\beta}[\mathscr{A}^{\alpha}{}_{\lambda\rho},a^{\rho\gamma}]-[\mathscr{A}^{\alpha}{}_{\rho\gamma},H_{\lambda\alpha\beta}]a^{\rho\gamma}+\nonumber\\
-[\mathscr{A}^{\alpha}{}_{\gamma\alpha},H_{\lambda\rho\beta}]a^{\rho\gamma}-\mathscr{A}^{\alpha}{}_{\beta\gamma}[H_{\lambda\rho\alpha},a^{\rho\gamma}]\bigg\}\delta
a^{\lambda\beta}\;.
\end{eqnarray}
At this point we introduce the operator $P$ defined as
\begin{equation}\label{57}
P_{\gamma}H_{\lambda\rho\beta}=\mathcal{D}H_{\lambda\rho\beta}+[\mathscr{A}^{\alpha}{}_{\lambda\gamma},H_{\alpha\rho\beta}]+[\mathscr{A}^{\alpha}{}_{\rho\gamma},H_{\lambda\alpha\beta}]+[\mathscr{A}^{\alpha}{}_{\beta\gamma},H_{\lambda\rho\alpha}]+[\mathscr{A}^{\alpha}{}_{\gamma\alpha},H_{\lambda\rho\beta}]\;.
\end{equation}
By using the last definition in (\ref{56}) one obtains
\begin{eqnarray}\label{58}
-\frac{1}{2}C^{\mu\nu}{}_{\alpha}\delta\mathscr{A}^{\alpha}{}_{\mu\nu}+\frac{1}{2}D^{\mu}\delta\mathscr{A}^{\alpha}{}_{\mu\alpha}=\frac{1}{4}\bigg\{2\mathscr{A}^{\alpha}{}_{[\beta\rho]}a^{\rho\gamma}H_{\alpha\lambda\gamma}+2H_{\alpha\lambda\beta}a^{\rho\gamma}\mathscr{A}^{\alpha}{}_{[\rho\lambda]}+\nonumber\\
-(P_{\gamma}H_{\lambda\rho\beta})a^{\rho\gamma}+[\mathscr{A}^{\alpha}{}_{\beta\gamma},H_{\lambda\rho\alpha}]a^{\rho\gamma}-H_{\alpha\gamma\beta}[\mathscr{A}^{\alpha}{}_{\lambda\rho},a^{\rho\gamma}]-\mathscr{A}^{\alpha}{}_{\beta\gamma}[H_{\lambda\rho\alpha},a^{\rho\gamma}]\bigg\}\delta
a^{\lambda\beta}\;.\nonumber\\
\;
\end{eqnarray}

We finally have all the ingredients that we need in order to write
the dynamical equations of the theory. Now we only have to find an
expression for the variation $\delta\rho$. The definition of
$\rho$ is given in (\ref{20}), and its variation can be
straightforwardly evaluated as follows
\begin{equation}\label{59}
\delta\rho=-\int\limits_{\RR^n}
\frac{d\xi}{\pi^{\frac{n}{2}}}\int_{0}^{1}ds\;
e^{-(1-s)A(\xi)}\delta a^{\mu\nu}\xi_{\mu}\xi_{\nu}
e^{-sA(\xi)}\;,
\end{equation}
where \be A(\xi)=a^{\mu\nu}\xi_\mu\xi_\nu\,. \ee Once we have the
expression (\ref{59}) for the variation, we can use the cyclic
property of the trace to write that
\begin{equation}\label{60}
\textrm{Tr}_{\;V}(\delta\rho\;\mathcal{R}) = \textrm{Tr}_{\;V}
\left[-\int\limits_{\RR^n}
\frac{d\xi}{\pi^{\frac{n}{2}}}\int_{0}^{1}ds\;
e^{-sA(\xi)}\mathcal{R}e^{-(1-s)A(\xi)}\xi_{\mu}\xi_{\nu}\right]\delta
a^{\mu\nu}\,.
\end{equation}

By combining (\ref{60}), (\ref{58}) and (\ref{32}) we obtain the
{\it noncommutative Einstein equations} in absence of matter,
namely
\begin{equation}\label{60a}
\mathcal{G}_{\mu\nu}=0\;,
\end{equation}
where
\begin{eqnarray}\label{61}
\mathcal{G}_{\mu\nu}&=& \frac{1}{2}\{\rho,\mathcal{R}_{\mu\nu}\}
+\mathcal{F}_{\mu\nu}+\frac{1}{2}\mathscr{A}^{\alpha}{}_{[\mu\rho]}
a^{\rho\gamma}H_{\alpha\nu\gamma}
+\frac{1}{2}H_{\alpha\lambda\nu}a^{\rho\gamma}
\mathscr{A}^{\alpha}{}_{[\rho\mu]}
-\frac{1}{4}(P_{\gamma}H_{\mu\rho\nu})a^{\rho\gamma}
+\nonumber\\
&+&\frac{1}{4}[\mathscr{A}^{\alpha}{}_{\nu\gamma},
H_{\mu\rho\alpha}]a^{\rho\gamma} -\frac{1}{4}H_{\alpha\gamma\nu}
[\mathscr{A}^{\alpha}{}_{\mu\rho},a^{\rho\gamma}]
-\frac{1}{4}\mathscr{A}^{\alpha}{}_{\nu\gamma}
[H_{\mu\rho\alpha},a^{\rho\gamma}]\;,
\end{eqnarray}
is the {\it noncommutative Einstein tensor},
$\mathcal{F}_{\mu\nu}$ is defined by \be \mathcal{F}_{\mu\nu} =
-\int\limits_{\RR^n}
\frac{d\xi}{\pi^{\frac{n}{2}}}\int_{0}^{1}ds\;
e^{-sA(\xi)}\mathcal{R}e^{-(1-s)A(\xi)}\xi_{\mu}\xi_{\nu}, \ee and
the tensor density $H$ has the explicit form
\begin{equation}\label{62}
H_{\alpha\lambda\gamma}=b_{\alpha\nu}(\delta^{\nu}{}_{\lambda}D^{\mu}-C^{\mu\nu}{}_{\lambda})b_{\mu\gamma}-b_{\lambda\nu}(\delta^{\nu}{}_{\gamma}D^{\mu}-C^{\mu\nu}{}_{\gamma})b_{\mu\alpha}-b_{\gamma\nu}(\delta^{\nu}{}_{\alpha}D^{\mu}-C^{\mu\nu}{}_{\alpha})b_{\mu\lambda}\;.
\end{equation}

These equations are the {\it main result of the present paper}.
One can show that the first two terms in the equations (\ref{61})
represent a straightforward generalization of Einstein's equation
to endomorphism-valued objects and the rest of the terms can be
considered as a genuine noncommutative part which is not present
in Einstein's equation. It is interesting to note that the pure
noncommutative part is completely described by the tensor density
$H_{\mu\nu\rho}$ defined in (\ref{62}).

Moreover the equation (\ref{60a}) satisfies the requirement
(\ref{26}), which, in words, expresses the necessity that our
model reduces, in the commutative limit, to the standard theory of
general relativity. In fact, the trace of the pure noncommutative
terms vanishes, because of the presence of the commutators, and
the first two terms just give
\begin{equation}\label{62a}
\frac{1}{N}\textrm{Tr}_{V}\left(\frac{1}{2}\{\rho,\mathcal{R}_{\mu\nu}\}+\mathcal{F}_{\mu\nu}\right)=\sqrt{g}\left(R_{\mu\nu}-\frac{1}{2}g_{\mu\nu}R\right)\;.
\end{equation}

For an arbitrary matrix algebra the equation (\ref{60a}) becomes
more complicated than the ordinary Einstein's equation due to
presence of the new tensor density $H_{\mu\nu\rho}$. We mention,
now, a particular case in which (\ref{60a}) simplifies. The
formalism used so far deals with geometric quantities which are
endomorphism-valued, namely they take values in $\textrm{End}(V)$.
By choosing a basis in the vector space $V$ we can represent
$\textrm{End}(V)$ by means of matrices. Let us suppose that the
algebra under consideration is Abelian, in this case all the
elements commute with each other and the tensor density
$H_{\mu\nu\rho}$ vanishes identically and the equation (\ref{60a})
becomes
\begin{equation}\label{62b}
\mathcal{R}_{\mu\nu}-\frac{1}{2}b_{\mu\nu}\mathcal{R}=0\;.
\end{equation}
Therefore, in case of a commutative matrix algebra, the equation
of motion of our model have the same form as Einstein's equation,
with the only difference that (\ref{62b}) is matrix-valued.

\section{The Action for the Matter Field}

In order to have a complete theory for the gravitational field we
need to describe the dynamics of the matter field in the framework
of matrix general relativity. The main idea is to extend the
general results of classical field theory. We will consider, in
the following, the dynamics of a multiplet of free scalar fields
propagating on a manifold $M$. We can construct an invariant
action by using the matrix valued metric $a^{\mu\nu}$ and the
measure $\rho$. A typical action is \be
S_{\textrm{matter}}(a,\varphi) =\frac{1}{4}\int_M
dx\;\left\{-\left<\partial_\mu\varphi, \{\rho, a^{\mu\nu}\}
\partial_\nu\varphi\right>
-\left<\varphi,\{\rho, Q\}\varphi\right> \right\}\,, \label{68}
\ee where $\left<\;,\;\right>$ denotes the fiber inner product on
the vector bundle $V$, and $Q$ is a constant mass matrix
determining the masses of the scalar fields. The equations of
motion of the scalar fields are then obviously \be
\left[-\partial_\mu\{\rho, a^{\mu\nu}\}\partial_\nu+\{\rho,Q\}
\right]\varphi=0\,.
\ee

The complete action of the gravity and matter is described then by
\begin{equation}\label{70a}
{S}(a,\varphi)=S_{\textrm{MGR}}(a)
+S_{\textrm{matter}}(a,\varphi)\;.
\end{equation}
By varying the above action with respect to $a^{\mu\nu}$ one
obtains the noncommutative Einstein equation in presence of matter
\be \mathcal{G}_{\mu\nu}=8\pi G N {\cal T}_{\mu\nu}\,, \ee where
${\cal T}_{\mu\nu}$ is the matrix energy-momentum tensor defined
by
\begin{equation}\label{66}
{\cal T}_{\mu\nu}=-\frac{1}{2}\frac{\delta
S_{\textrm{matter}}}{\delta a^{\mu\nu}}\;.
\end{equation}
By using the explicit lagrangian (\ref{68}) for the matter field,
we obtain the expression for the energy-momentum tensor
\begin{equation}\label{70}
{\cal
T}_{\mu\nu}=\frac{1}{8}\left[\{\rho,\partial_{\mu}\varphi\otimes\partial_{\nu}\varphi\}+\mathcal{M}_{\mu\nu}+\mathcal{N}_{\mu\nu}\right]+(\mu\leftrightarrow\nu)\;,
\end{equation}
where the explicit form of $\mathcal{M}_{\mu\nu}$ and
$\mathcal{N}_{\mu\nu}$ is obtained by using the variation of the
scalar density $\rho$ in (\ref{59}), namely
\begin{equation}\label{71}
\mathcal{M}_{\mu\nu}= -\int\limits_{\RR^n}
\frac{d\xi}{\pi^{\frac{n}{2}}}\int_{0}^{1}ds\;
e^{-sA(\xi)}\{a^{\alpha\beta},\partial_{\alpha}\varphi\otimes\partial_{\beta}\varphi\}e^{-(1-s)A(\xi)}\xi_{\mu}\xi_{\nu}\;,
\end{equation}
and
\begin{equation}\label{72}
\mathcal{N}_{\mu\nu}=-\int\limits_{\RR^n}
\frac{d\xi}{\pi^{\frac{n}{2}}}\int_{0}^{1}ds\;
e^{-sA(\xi)}\{Q,\varphi\otimes\varphi\}e^{-(1-s)A(\xi)}\xi_{\mu}\xi_{\nu}\;.
\end{equation}

It is worth remarking, here, that the above formula (\ref{70}) for
the energy-momentum tensor ${\cal T}_{\mu\nu}$ reduces, in the
commutative limit, to the standard result, e.g. \cite{FU}.

\section{Conclusions}

The main idea of this new model is to describe the gravitational
field by a multiplet of gauge fields with some internal structure.
For this purpose the metric field $g^{\mu\nu}$, which describes
gravity in general relativity, is replaced by a matrix-valued
2-tensor field $a^{\mu\nu}$. This allows the model to have a much
richer content in describing gravitational phenomena. A more
general geometric picture is developed by allowing the metric to
be matrix-valued. Most of the geometric quantities, used in
describing gravity, can be generalized to be endomorphism-valued.
In this framework it is possible to introduce an action for the
gravitational field which is diffeomorphisms and gauge invariant,
that leads, after performing the variation with respect to
$a^{\mu\nu}$, to the modified (noncommutative) Einstein equation.
It is interesting that the noncommutative part of the modified
equations only depends on a specific tensor density
$H_{\mu\nu\rho}$ and on a linear combination of its commutators.

With an explicit expression for the noncommutative Einstein
equation, it will be possible to study some particular simple
solutions of (\ref{60a}); for example a static and spherically
symmetric solution, which would describe the gravitational field
outside a spherically symmetric massive body, or a spherically
symmetric homogeneous solution, that would describe noncommutative
cosmological models. These simple examples are very interesting
and we plan to study them systematically elsewhere.

We would like to make a final remark. In our model all the
geometric quantities that we need to develop the formalism are
endomorphism-valued. Once a basis for the vector bundle $V$ has
been fixed, we can represent elements of $\textrm{End}(V)$ by
matrices. Of course the description of physical phenomena has to
be independent from the particular realization of the
representation. This is, ultimately, related to the gauge
invariance of the theory. We believe that by an opportune choice
of gauge, namely an opportune representation of $\textrm{End}(V)$
by matrices, the dynamical equation (\ref{60a}) could be
simplified further. The search for such particular gauge, if it
exists, requires further studies in matrix differential geometry
and matrix general relativity.

\end{document}